\documentclass[preprint,12pt,authoryear]{elsarticle}
\journal{Journal of the Mechanics and Physics of Solids}
\usepackage{graphicx}
\usepackage{amssymb,amstext,amsmath}
\usepackage{natbib}
\usepackage{booktabs}
\usepackage{physics}

\begin{document}
\begin{frontmatter}
\title{A phase field model of coupled crack and dislocations: emission, blunting, and the necessity of dissipative toughening}
\author[1,2]{Khanh Chau Le\corref{cor}}
\ead{lekhanhchau@tdtu.edu.vn}
\author[1,2]{Thi My Kieu Tran}
\cortext[cor]{Corresponding author}
\address[1]{Mechanics of Advanced Materials and Structures, Institute for Advanced Study in Technology, Ton Duc Thang University, Ho Chi Minh City, Vietnam}
\address[2]{Faculty of Civil Engineering, Ton Duc Thang University, Ho Chi Minh City, Vietnam}
\begin{abstract}
We propose a phase field model of a macrocracked single crystal in which the crack and the geometrically necessary dislocations descend from a single energy functional. Energy minimization alone then decides dislocation nucleation, through an integral criterion evaluated in closed form along slip chords. The criterion yields a size effect inaccessible to point-wise strength conditions: a grain-size-dependent yield stress. With slip suppressed the model reproduces Griffith fracture; with fracture suppressed, the nucleation load measured by the full non-smooth solver agrees with the closed-form nucleation criterion to four percent. The coupled computations produce a two-stage response: at loads an order of magnitude below cleavage, dislocation bands emitted from the notch tip blunt and shield it, raising the initiation load; once the crack grows, the bands heal; a compact cluster of like-signed dislocations travels with the tip, its canceling partner walls pinned at the grain boundary, and the dissipated fracture resistance equals the elastic one. In the purely energetic, dissipationless limit, emission shields the crack but does not toughen it; toughening requires dissipation, incorporated in the sequel through the threshold resistance to dislocation motion.
\end{abstract}

\end{frontmatter}

\section{Introduction}
\label{sec:intro}

Whether a stressed crack in a crystal advances by cleavage or is blunted by the emission of dislocations is the oldest question of the fracture mechanics of metals \citep{rice1974ductile,gumbsch1998controlling}, and it is a question about \emph{nucleation}: whether, at a given load, the crystal prefers to create new surface or new dislocations. Discrete dislocation plasticity coupled to a cohesive description of decohesion \citep{cleveringa2000discrete} resolves this competition dislocation by dislocation, but with nucleation as a constitutive input and dissipation inseparable from the dynamics: it answers what fracture costs while postulating what nucleates. The variational phase field approach to fracture \citep{francfort1998revisiting,bourdin2000numerical,ambrosio1990approximation} answers the cleavage half of the alternative by energy minimization, and its success has motivated numerous extensions to ductile fracture in which the damage field is coupled to plasticity.

These couplings now form three families. In the crystal-plasticity framework of DAMASK \citep{shanthraj2016phase,shanthraj2017elasto,roters2019damask}, an elasto-viscoplastic flow rule with dislocation densities as internal variables drives the damage field through the plastic work. In gradient-extended couplings, the plastic variable itself carries spatial gradients: micromorphic formulations regularize the interaction of shear bands with the crack \citep{miehe2016aphase,miehe2016bphase}, and mechanism-based strain gradient plasticity feeds the geometrically necessary dislocations, through Taylor hardening, into the flow stress, capturing its elevation ahead of the tip \citep{martinez2016strain,kristensen2020phase}. In the recent nucleation-oriented models, finally, damage onset is calibrated to a strength surface---through tailored energy decompositions \citep{de2022nucleation,zolesi2024stability} or, in the most recent formulation, through a reversible eigenstrain endowed with a degree-one strength potential \citep{vicentini2025variational}.

In the first two families, dislocation activity is governed by a flow rule---a point-wise condition on the stress, satisfied wherever it is met, and dissipative by construction---and the consequences are systematic. Dislocation nucleation is not an event: plastic flow exists wherever the criterion is met, with no threshold load, no emission site, and no discrete dislocation content. Where a plastic length is present, it enters the hardening or the regularization---it elevates the flow stress or sets the band width---but it does not produce a nucleation threshold: the size effects it generates are of hardening type, and the $d^{-1}$ dependence of the initial yield stress of small grains, derived in Section~\ref{sec:crit-shear}, is not among them. Blunting has no mechanism: plastic strain smears the tip fields but produces no slip step on the crack surface. And because every increment of slip dissipates, the energetic content of dislocation emission---what shielding costs, as distinct from what wake dissipation costs---cannot be isolated within these formulations; this question is posed and answered in Section~\ref{sec:coupled-null}. The last of the nucleation-oriented models \citep{vicentini2025variational} is, instructively, closer in structure to the present model: its eigenstrain is reversible, and its strength potential is positively homogeneous of degree one. The difference lies in where thermodynamics permits the degree-one term to act. A free energy may depend only on state variables. The cohesive eigenstrain of these models is one---a reversible measure of the crack opening---and a degree-one energy in the variable itself is then admissible; it yields a point-wise threshold, a strength surface, containing no length. Crystallographic slip, by contrast, is not a state variable: the same lattice configuration is reached by arbitrarily many slip histories, and slip therefore cannot carry free energy \citep{berdichevsky2006continuum,berdichevsky2006thermodynamics}. What is a state variable is the incompatible part of its gradient---the dislocation density \citep{nye1953some,bilby1955types,kroner1955fundamentale}---and the degree-one term must accordingly act on the gradient. This placement, dictated by thermodynamics rather than chosen, is what converts the point-wise threshold into an integral one, evaluated along the characteristics, and endows it with the size and geometry dependence that a strength surface cannot express. Moreover, the eigenstrain of these models represents the opening of a cohesive crack, not crystallographic slip: they address the nucleation of fracture, not of dislocations.

The regularization length of the fracture energy has, by contrast, earned a physical reading through its calibration against the material strength; but it regularizes the crack and says nothing about the plasticity. In the model proposed here the plastic side carries its own, independently identified lengths: the dislocation length 
$\ell_d = 1/(b\rho_s)$, set by the Burgers vector and the saturated dislocation density, and the threshold modulus $q_c = \mu k/(b\rho_s)$ of the dislocation energy.
 
The model combines three classical ingredients. From the variational
theory of fracture it takes the Ambrosio--Tortorelli representation of
the crack by a damage field. From the physics of the crack tip it
takes the Rice--Thomson picture of emission, blunting, and shielding
\citep{rice1974ductile}. And from the continuum dislocation theory of
single crystals
\citep{berdichevsky2006continuum,berdichevsky2006thermodynamics,
berdichevsky2007dislocation,le2008analytical} it takes the energy of
the geometrically necessary dislocations
\citep{nye1953some,bilby1955types,kroner1955fundamentale} as a
logarithmic function of their density---whose expansion about the
dislocation-free state begins with a term of \emph{degree one}. This
nonsmoothness is not a technical inconvenience but the physical heart
of the model: a finite energetic cost per unit dislocation content at
vanishing content is precisely what endows the theory with a
nucleation \emph{threshold}, in the same way that the degree-one
dissipation of dry friction endows a block with a sticking threshold.
To our knowledge, the resulting formulation is the first variational
phase field model of a crack coupled to dislocations in which
nucleation, blunting, and crack growth all emerge from the
minimization of one energy functional.
 
The contributions of the paper are the following.
\begin{enumerate}
\item A dislocation nucleation criterion of integral type, derived as
the stability condition of the dislocation-free state
(Section~3): the crystal remains dislocation-free as long as the work
extractable by any trial slip is outweighed by its dislocation-energy
cost. By convex duality the criterion is evaluated in closed form
along the slip chords, and it contains the size effects that
point-wise conditions cannot produce: the inverse-thickness threshold
of constrained shear, a grain-size-dependent yield stress consistent
with the $d^{-1}$ scaling reported at small grain sizes
\citep{li1963petch,conrad1963effect,dunstan2014grain}, and a vanishing threshold on slip planes grazing bare free surfaces, which thereby act as
dislocation sources, while passivated surfaces and grain boundaries
restore a finite threshold.
\item A numerical method (Section~4) that treats the non-smoothness
exactly---an alternating-direction splitting for the slip and a
primal--dual active set method for the irreversible damage---together
with two verification instruments that appear to be new in this
context: the equipartition of the fracture energy as a mesh-adequacy
diagnostic, and an a posteriori energy bound on the constraint gap as
the correct convergence certificate for minimizers containing
measure-valued slip.
\item A validation of both limits of the model (Section~5): with slip suppressed, regularized Griffith fracture is recovered with an effective toughness converging to $G_c$ under mesh refinement; with fracture suppressed, the nucleation load measured by the full non-smooth solver agrees with the closed-form criterion to four
percent, including the kinetics of the onset.
\item The coupled computations divide the response into two stages: at loads an order of magnitude below the cleavage load, dislocations are emitted in bands from the notch tip and blunt it, and the shielding they provide raises the initiation load by a factor 1.2--1.7 in stress intensity. Once the crack grows, the band structure heals; a compact cluster of like-signed dislocations travels with the tip at a standoff of the order of the damage length, its partner walls pinned at the grain boundary; the slope of the fracture energy with respect to crack advance equals the elastic resistance to a fraction of a percent, the surviving dislocation energy being stored, not dissipated. In the purely energetic, dissipationless limit, emission shields the crack---it delays initiation and screens the growing tip---but does not toughen the crystal: the classical distinction between shielding and toughening
\citep{ritchie2011conflicts}, here derived rather than postulated.
\end{enumerate}
 
The last result fixes the program of which this paper is the first
stage. The model is predictive in the strict sense---its thresholds
are computed before the simulations that confirm them, and its
observables (emission loads, standoff, size-effect scalings, the
event sequence) are falsifiable---but it is deliberately athermal
and dissipationless, and the null result demonstrates that this limit,
while it decides what nucleates and what structures exist, cannot
decide what fracture costs. Toughening is dissipation in the wake,
and its incorporation through the threshold resistance to dislocation
motion, together with the redundant dislocations and the
configurational temperature of the thermodynamic dislocation theory,
is the subject of the sequel; the thermally activated kinetics that
turn the competition between emission and cleavage into a
temperature-dependent one---the brittle-to-ductile transition---is
the stage after that. That final stage has both an experimental destination and a theoretical counterpart already in place. The destination is the transition measured
on tungsten single crystals by \citet{gumbsch1998controlling}, whose
loading-rate dependence ties the transition to thermally activated
dislocation processes at the tip and whose separation of the roles of
nucleation and mobility is the experimental form of the question posed
here. The counterpart is the theory of the transition developed within
the thermodynamic dislocation framework \citep{langer2021fracture,le2023theory},
which treats the crack and its dislocation population as a whole---a
mean-field dynamical system whose kinetics render the transition
temperature computable, but into which the spatial mechanics of
emission, its energetic cost, and the shielding it provides enter as
inputs. The present model is built at the opposite end: it resolves the
fields and computes precisely those inputs, and the staged program is
what connects the two. 

The paper is organized as follows. Section~2
formulates the model; Section~3 derives the nucleation criterion and
its consequences; Section~4 describes the numerical method and its
certificates; Section~5 validates the two limits; Section~6 presents
the coupled results; Section~7 discusses the findings and the
outlook.
 
\section{Variational formulation}
\label{sec:model}


\subsection{Kinematics and fields}
\label{sec:model-kin}

Consider a single crystal occupying in the undeformed state the 2D domain $\Omega$. Under the plane strain loading condition the crystal deforms with strains assumed small everywhere so that the geometrically linear theory applies. The state of the crystal is described by three fields: the 2D displacement vector $\vb{u}$, with total (small) strain $\vb*{\varepsilon}(\vb{u})=\tfrac12(\grad\vb{u}+\grad\vb{u}^{\!\top})$; the scalar plastic slip $\beta$ on a single slip system with slip direction $\vb{s}=(\cos\varphi,\sin\varphi)$ and slip-plane normal $\vb{m}=(-\sin\varphi,\cos\varphi)$; and the scalar damage field $\alpha\in[0,1]$, with $\alpha=0$ intact and $\alpha=1$ fully broken material. The plastic strain is
\begin{equation}
\vb*{\varepsilon}^p = \beta\,\vb{P},
\quad
\vb{P}=\tfrac12\big(\vb{s}\otimes\vb{m}+\vb{m}\otimes\vb{s}\big),
\label{eq:epsp}
\end{equation}
and the elastic strain is
$\vb*{\varepsilon}^e=\vb*{\varepsilon}(\vb{u})-\beta\vb{P}$.

Gradients of slip are of two kinds. A slip field varying only in the
direction of $\vb{m}$, i.e.\ from slip plane to slip plane, is
rank-one compatible and stores no defects. A gradient \emph{along the
slip direction}, $\partial_s\beta=\vb{s}\vdot\grad\beta$, corresponding
to slip terminating within its own plane, measures the resultant Burgers vector of all geometrically necessary edge dislocations, whose dislocation lines are perpendicular to the plane under consideration \citep{nye1953some,bilby1955types,kroner1955fundamentale}. So, its (signed) density reads
\begin{equation}
\rho = \frac{1}{b}\,\partial_s\beta,
\label{eq:rho}
\end{equation}
where $b$ is the magnitude of the Burgers vector. The straight lines parallel to $\vb{s}$ are thus the characteristics of the defect content; their segments inside $\Omega$, the \emph{slip chords}, organize the entire nucleation analysis of
Section~\ref{sec:criterion}.

\subsection{Energy functional}
\label{sec:model-energy}

The model is defined by a single energy functional per unit depth,
\begin{equation}
J[\vb{u},\beta,\alpha]
=
\int_\Omega
\Big[\,
g(\alpha)\,\psi_e\big(\vb*{\varepsilon}(\vb{u})-\beta\vb{P}\big)
+\psi_d\big(\partial_s\beta\big)
+\psi_c\big(\alpha,\grad\alpha\big)
\,\Big]\dd{a},
\label{eq:J}
\end{equation}
with $\dd{a}=\dd{x} \dd{y}$ being the area element, whose three densities carry the elastic, dislocation, and fracture physics respectively. We postulate the following variational principle: The true fields $\check{\vb{u}}$, $\check{\beta}$, and $\check{\alpha}$ minimize functional \eqref{eq:J} among all kinematically admissible fields. Below are the descriptions of the three energy contributions.

\emph{Elastic energy.} $\psi_e(\varepsilon^e)=\tfrac12\,
\vb*{\varepsilon}^e\boldsymbol{:}\mathbb{C}\boldsymbol{:}\vb*{\varepsilon}^e$ with the isotropic 4-rank plane-strain stiffness tensor $\mathbb{C}$ of Lam\'e constants
$(\lambda,\mu)$, degraded by the standard quadratic function
\begin{equation}
g(\alpha)=(1-\eta)\,(1-\alpha)^2+\eta,
\quad 0<\eta\ll 1,
\label{eq:g}
\end{equation}
with $\eta$ a small residual stiffness. Only the elastic energy is degraded: damage is driven by the stored elastic energy alone, as in the Rice--Thomson setting \citep{rice1974ductile}, while the dislocation energy below is a property of the crystal lattice and survives in the wake of the crack.

\emph{Dislocation energy.} Following the continuum dislocation theory
of single crystals having one active slip system \citep{berdichevsky2006thermodynamics,berdichevsky2006continuum}, the energy of the geometrically necessary dislocations is the logarithmic function of the density
\eqref{eq:rho}, of which we retain the first two terms of the expansion about $\rho=0$,
\begin{equation}
\psi_d
=
\mu k\left[\frac{\abs{\rho}}{\rho_s}
+\frac{1}{2}\frac{\rho^2}{\rho_s^2}\right]
=
q_c\,\abs{\partial_s\beta}
+\frac{c_2}{2}\,\big(\partial_s\beta\big)^2 ,
\label{eq:psid}
\end{equation}
where $\rho_s$ is the saturated dislocation density, $k$ a
dimensionless energy factor of order $10^{-3}$--$10^{-1}$, and
\begin{equation}
\ell_d = \frac{1}{b\rho_s},
\quad
q_c = \mu k\,\ell_d,
\quad
c_2 = \mu k\,\ell_d^{\,2}.
\label{eq:qc}
\end{equation}
The degree-one term makes $\psi_d$ nonsmooth at $\rho=0$: dislocation
content has a finite energetic cost per unit density at vanishing density, which is the origin of the nucleation \emph{threshold} derived in Section~\ref{sec:criterion}. The quadratic term regularizes dislocation walls to the thickness scale $c_2/q_c=\ell_d$; it causes kinematic hardening but plays no role in dislocation nucleation.

\emph{Fracture energy.} Following the variational approach to brittle fracture \citep{francfort1998revisiting,bourdin2000numerical}, the crack is represented not as a surface of displacement discontinuity as in the traditional fracture mechanics \citep{griffith1920phenomena,rice1968path,stumpf1990variational} but through the
damage field $\alpha$, which localizes in narrow bands whose width is set by a regularization length $\ell$; the Griffith surface energy is recovered in the limit $\ell\to 0$ in the sense of $\Gamma$-convergence, by the elliptic regularization of
\citet{ambrosio1990approximation} (AT2). We adopt this AT2-regularization in the
form
\begin{equation}
\psi_c = G_c\left(\frac{\alpha^2}{4\ell}
+\ell\,\abs{\grad\alpha}^2\right),
\label{eq:psic}
\end{equation}
with fracture toughness $G_c$ and regularization length $\ell$, so
that a fully developed damage band of profile
$\alpha=e^{-\abs{d}/2\ell}$ carries surface energy exactly $G_c$ per
unit length.

Three material lengths therefore coexist---the damage length $\ell$,
the dislocation length $\ell_d$, and (below) the notch root radius---and, in contrast to formulations in which a single length is
calibrated to regularize the crack, each carries independent physical
identity: $\ell$ from $(G_c,\sigma_c)$, $\ell_d$ from $(b,\rho_s)$.

\emph{Irreversibility.} Damage is irreversible: along the loading
program, $\alpha$ is constrained to be point-wise nondecreasing, and
$\alpha\le 1$. The slip $\beta$ carries \emph{no} irreversibility
constraint: the present paper studies the purely energetic
(dissipationless) limit of the theory, in which plastic slip may heal.
This modeling choice is deliberate; its principal consequence---that
dislocation emission shields the crack but produces exactly zero
steady-state toughening---is derived in Section~\ref{sec:coupled} and motivates the
dissipative extension of the theory.

\subsection{Geometry, loading, and boundary conditions}
\label{sec:model-bc}
\begin{figure}[htb]
    \centering
    \includegraphics[width=5cm]{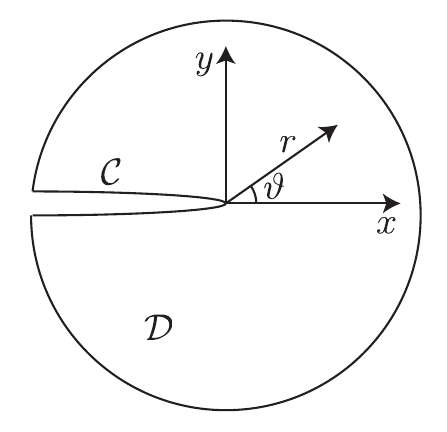}
    \caption{A notched disk.}
    \label{fig:1}
\end{figure}
The primary configuration is a macrocracked single crystal modeled as the notched disk (see Fig.~\ref{fig:1})
\begin{equation}
\Omega=\mathcal{D}\setminus\mathcal{C},
\quad
\mathcal{D}:\ x^2+y^2\le 1,
\quad
\mathcal{C}:\ (x+1)^2+\frac{y^2}{\epsilon^2}\le 1,
\label{eq:domain}
\end{equation}
with lengths measured in units of the specimen size $L_0$ (one micron in the
calibration below). The ellipse $\mathcal{C}$ is a rounded model of the pre-existing crack: its tip lies at the origin with root radius $\rho_{\rm tip}=\epsilon^2L_0$, so that $\epsilon$ interpolates between a blunt notch ($\rho_{\rm tip}\gtrsim\ell$) and a mathematically sharp crack ($\rho_{\rm tip}\ll\ell$), in which case the damage length takes over the regularization of the tip. As a whole, this notched disk models the grain surrounding the tip of a macrocrack in a polycrystalline material.

On the outer circle $\partial\mathcal{D}$ the displacement of the plane-strain mode~I crack-tip field is prescribed,
\begin{equation}
\begin{split}
u_1=\kappa\sqrt{\frac{r}{2\pi}}\,\cos\frac{\vartheta}{2}
\Big(1-2\nu+\sin^2\frac{\vartheta}{2}\Big),
\\
u_2=\kappa\sqrt{\frac{r}{2\pi}}\,\sin\frac{\vartheta}{2}
\Big(2-2\nu-\cos^2\frac{\vartheta}{2}\Big),
\end{split}
\label{eq:Kfield}
\end{equation}
in polar coordinates $(r,\vartheta)$ centered at the crack tip, with
$\nu$ the Poisson ratio. The amplitude $\kappa$ is the normalized
stress intensity factor; comparison with the standard form of linear fracture mechanics shows $\kappa=K_I/\mu$, so that the Griffith load of the sharp crack is
$\kappa_c=\sqrt{E'G_c}/\mu$ with $E'=2\mu/(1-\nu)$. In the surfing
protocol of Section~\ref{sec:num-load} the center of the field \eqref{eq:Kfield} is
translated along the crack path at fixed $\kappa$. On
$\partial\mathcal{D}$ we further impose $\alpha=0$; the notch surface
$\partial\mathcal{C}$ is traction-free with the natural condition for
$\alpha$.

For the slip, the boundary $\partial\Omega$ is partitioned into a \emph{passivated} part $\Gamma_p$, on which $\beta=0$ (a coated/oxidized surface or a grain boundary through which dislocations cannot pass, so as a rule $\partial \mathcal{D}\subseteq \Gamma_p$), and a \emph{free} part $\Gamma_f$, on which $\beta$ satisfies the natural condition (a bare surface through which dislocations enter and leave, so as a rule $\partial \mathcal{C}\subseteq \Gamma_f$). This partition is a physical modeling choice, not a technicality: as shown in Section~\ref{sec:crit-surface}, it decides whether the crystal possesses a finite nucleation threshold at all. Both configurations are studied below. Note that, in case of crack growth/propagation, the Griffith cleavage dominates over the dislocation activity, so the boundary condition for $\beta$ posed on the translated surface does not affect the whole behavior of the crack. 

\subsection{Parameter identification}
\label{sec:model-param}

The stress unit is $\mu$ and the length unit the specimen size $L_0$ (one micron: the grain surrounding the macrocrack tip). The calibration is to tungsten at the micron scale: $\mu=161$~GPa, $b=0.274$~nm. The saturated density is taken as $\rho_s=1.8 \times 10^{17}\,\mathrm{m}^{-2}$, giving the dislocation length $\ell_d=1/(b\rho_s)=20$~nm. The dimensionless factor $k$ is the small-density coefficient of the logarithmic dislocation energy, and only the product $q_c=\mu k/(b\rho_s)$ enters the nucleation thresholds. The fracture parameters $G_c=3.2\times 10^{-4} \mu L_0=51.5$~J/m$^2$ correspond to a cleavage toughness $K_{Ic}=4.9\,\mathrm{MPa}\sqrt{\mathrm{m}}$, of the order of the low-temperature single-crystal values. The parameters of each computation are collected in Table~\ref{tab:calibration}; where they depart from this calibration---notably the factor $k$ of the nucleation test of Section~\ref{sec:val-nucl}---the departure is a validation setting chosen for the sharpness of the test, and is flagged there. 

\begin{table}[t]
\centering
\caption{Parameters of the three computational configurations. Code
units: lengths in units of $L_0 = 1\,\mu$m, stresses in units of
$\mu$. Physical values of the coupled configuration (tungsten,
$\mu = 161$\,GPa, $b = 0.274$\,nm): $\ell = \ell_d = 20$\,nm,
$\rho_{\rm tip} = 2.5$\,nm,
$\rho_s = 1/(b\ell_d) = 1.8\times10^{17}\,$m$^{-2}$,
$G_c = 51.5$\,J/m$^2$, i.e.\ $K_{Ic} = 4.9$\,MPa$\sqrt{\rm m}$.}
\label{tab:calibration}
\footnotesize
\setlength{\tabcolsep}{5pt}
\begin{tabular}{lccc}
\toprule
 & Griffith & Nucleation & Coupled \\
 & (Sec.~\ref{sec:val-griffith}) & (Sec.~\ref{sec:val-nucl})
 & (Sec.~\ref{sec:coupled}) \\
\midrule
domain            & disk        & square            & disk \\
suppressed        & $\beta$     & $\alpha$ (frozen) & ---  \\
$\nu$             & $0.30$      & $0.30$            & $0.30$ \\
$G_c$             & $3.2\times10^{-4}$ & ---        & $3.2\times10^{-4}$ \\
$\ell$            & $0.02$      & $0.02$            & $0.02$ \\
$\epsilon$        & $0.05$      & --- $^{\rm (a)}$  & $0.05$ \\
$\varphi$         & ---         & $70.53^\circ$     & $70.53^\circ$ \\
$k$               & ---         & $0.42$ $^{\rm (b)}$ & $0.03$ $^{\rm (c)}$ \\
$\ell_d$          & ---         & $0.02$            & $0.02$ \\
$q_c=\mu k\ell_d$ & ---         & $8.4\times 10^{-3}$            & $6\times10^{-4}$ \\
$c_2=\mu k\ell_d^2$ & ---       & $1.68\times 10^{-4}$            & $1.2\times10^{-5}$ \\
$\beta$ b.c.      & ---         & $\beta=0$ on $\partial\Omega$
                                                    & $^{\rm (d)}$ \\
loading           & surfing $^{\rm (e)}$ & ramp $\bar v$
                                                    & two-phase $^{\rm (f)}$ \\
$h/\ell$ (path)   & $2.8$--$0.085$ & $200^2$ bilin. & $0.085$ \\
\bottomrule
\end{tabular}

\medskip
\raggedright\footnotesize
$^{\rm (a)}$ Seeded crack of half-length $a_0 = 0.15$ in place of the
notch.\;
$^{\rm (b)}$ Validation setting, not a calibration: chosen to place
the threshold inside the Rice--Thomson window of the specimen so that
case~(i) of the chord criterion is tested sharply.\;
$^{\rm (c)}$ Physical calibration; only $q_c = \mu k/(b\rho_s)$
enters the thresholds, while $\ell_d = c_2/q_c$ sets the wall
thickness.\;
$^{\rm (d)}$ $\beta = 0$ on $\partial\mathcal D$ (grain boundary),
natural on $\partial\mathcal C$ (bare notch).\;
$^{\rm (e)}$ $G/G_c = 1.2$--$1.69$.\;
$^{\rm (f)}$ Initiation at $G_{\rm init}/G_c \approx 3.0$, measurement
at $G_{\rm meas}/G_c = 1.5$ (\ref{app:certificates}).
\end{table} 

\section{The dislocation nucleation criterion}
\label{sec:criterion}


\subsection{The energetic criterion}
\label{sec:crit-primal}

We first ask under what conditions the dislocation-free state $\beta \equiv 0$
remains the minimizer of the energy at a given deformation and damage
field. Freeze $(\vb{u},\alpha)$ and regard the functional \eqref{eq:J} of
Section~\ref{sec:model} as a functional of the plastic slip alone,
\begin{equation}
E[\beta] = \int_\Omega \Big[\, g(\alpha)\,
\psi_e\big(\vb*{\varepsilon}(\vb{u})-\beta \vb{P}\big)
+ q_c\,\abs{\partial_s\beta}
+ \tfrac{c_2}{2}\,(\partial_s\beta)^2 \,\Big]\dd{a},
\label{eq:Ebeta}
\end{equation}
with $\vb{P}$, $q_c=\mu k\ell_d$, and $c_2=\mu k\ell_d^{\,2}$ as in
\eqref{eq:epsp} and \eqref{eq:qc}. Admissible slips satisfy
$\beta = 0$ on the passivated part $\Gamma_p$ of the boundary and are
unconstrained on the free part $\Gamma_f$.

Two structural facts fix the character of the nucleation condition.
First, $E$ is \emph{convex} in $\beta$: the elastic term is positive definite in $\beta$, the bounded-variation (BV) term $q_c\,\abs{\partial_s\beta}$ is convex and positively homogeneous of degree one, and the $c_2$-term is a convex quadratic in $\partial_s\beta$. Local stability of $\beta=0$ and global energetic optimality therefore coincide; there is no gap between a stability threshold and an energetic threshold, exactly as in the anti-plane and constrained-shear problems of continuum
dislocation theory \citep{berdichevsky2007dislocation,le2008analytical}. Second, the quadratic terms contribute nothing at
first order, so the threshold is governed by the competition between the
elastic driving term and the degree-one BV term alone; the modulus $c_2$
sets the thickness of dislocation walls \emph{after} nucleation but does
not enter the nucleation condition.

The first-variation (subdifferential) condition for $\beta = 0$ to
minimize \eqref{eq:Ebeta} reads: for every admissible trial slip
$\delta\beta$,
\begin{equation}
\Big|\int_\Omega \tau\,\delta\beta \dd{a}\Big|
\le
q_c \int_\Omega \abs{\partial_s\,\delta\beta}\dd{a},
\quad
\tau := \vb*{\sigma} \boldsymbol{:} \vb{P},
\quad
\vb*{\sigma}= g(\alpha)\,\mathbb{C}\boldsymbol{:} \vb*{\varepsilon}(\vb{u}),
\label{eq:crit}
\end{equation}
The left-hand side is the work the resolved shear stress (Schmid's stress) $\tau$ of the (damage-degraded) stress field can extract from the trial slip distribution; the right-hand side is the dislocation energy the trial must pay. Nucleation occurs when, for some admissible $\delta\beta$, extraction exceeds cost. We emphasize the \emph{integral, nonlocal} character of \eqref{eq:crit}, in the tradition of the energetic dislocation-nucleation thresholds of continuum dislocation theory \citep{berdichevsky2007dislocation,le2008analytical}: In contrast to phase field models of fracture with strength surface \citep{de2022nucleation,zolesi2024stability,vicentini2025variational} no point-wise yield condition $\tau \le \tau_c$ is postulated anywhere in the model, and none would reproduce the size and geometry dependence derived below.

\subsection{Dual form: the micro-stress and the chord evaluation}
\label{sec:crit-dual}

The operator $\partial_s$ differentiates only along the slip direction;
its characteristics are the straight lines
$\vb{x}(s) = \vb{x}_0 + s\,\vb{s}$, and the intersection of any such
line with $\Omega$ is a union of segments which we call \emph{slip
chords}. (In a multiply-notched domain a single line may contribute
several chords; each carries its own end conditions.) A chord end lying
on $\Gamma_p$ is called \emph{pinned}, one lying transversally on
$\Gamma_f$ \emph{free}. The supremum implicit in \eqref{eq:crit}
decomposes along this foliation, and a one-dimensional duality
computation on each chord yields the following evaluation.

\medskip
\noindent\textbf{Chord criterion.}
\emph{For a chord parametrized by arclength $s\in[0,L]$ define
\begin{equation}
Q(s) = \int_0^{s} \tau\big(\vb{x}(s')\big)\dd{s'} ,
\label{eq:Q}
\end{equation}
and let
$\operatorname{osc}_{[0,L]} Q := \max_{s\in[0,L]} Q(s)
- \min_{s\in[0,L]} Q(s)$
denote the oscillation of $Q$ over the chord.
Condition \eqref{eq:crit} holds if and only if on every chord there
exists a \emph{micro-stress} $q(s) = Q(s) - c$, with $c$ constant on the
chord, such that
\begin{equation}
\abs{q(s)} \,\le\, q_c \ \ \text{on } [0,L],
\quad
q = 0 \ \text{at every free end.}
\label{eq:qcond}
\end{equation}
Explicitly:}
\begin{itemize}
\item[\emph{(i)}] \emph{pinned--pinned chord: $c$ is unconstrained, and
\eqref{eq:qcond} is solvable if and only if}
\ $\operatorname{osc}_{[0,L]} Q \,\le\, 2 q_c$;
\item[\emph{(ii)}] \emph{pinned--free chord (free end at $s=L$, say):
$c = Q(L)$ is forced, i.e.\ $q(s) = -\int_s^L \tau \dd{s'}$, and the
condition is}\ $\max_s \abs{\int_s^L \tau \dd{s'}} \le q_c$;
\item[\emph{(iii)}] \emph{free--free chord: solvability requires the
chord balance $\int_0^L \tau \dd{s} = 0$, and then
$\max_s\abs{Q(s)} \le q_c$. If the balance fails, the threshold is
zero: the chord-constant slip $\delta\beta = \mathrm{const}$ costs no
dislocation energy and extracts work at any load.}
\end{itemize}

\emph{Proof sketch.} Fix a chord and restrict \eqref{eq:crit} to trial
fields supported in a thin tube around it; the transverse direction
contributes no BV cost, so the worst case is one-dimensional. For
$v(s)$ with $v=0$ at pinned ends and $\int_0^L\abs{v'}\dd{s}\le 1$,
integration by parts gives
$\int_0^L \tau v \dd{s} = -\int_0^L \big(Q-c\big)\,v'\dd{s}$ for any
$c$ compatible with the end conditions (a boundary term
$\big[(Q-c)v\big]$ survives at free ends and is disposed of by the
choice of $c$, or, on free--free chords, by the additional costless
constant mode). Taking the supremum over the unit ball of measures $v'$
yields $\min_c \max_s\abs{Q-c}$ subject to the stated end constraints,
which is the content of (i)--(iii). $\square$

\medskip
The duality between the boundary conditions of $\beta$ and of $q$---\emph{pinned slip $\leftrightarrow$ free micro-stress, free slip
$\leftrightarrow$ vanishing micro-stress}---is the transcription, at
the level of the nucleation problem, of the natural/essential pairing
of the underlying variational structure. The field $q$ is precisely the
Lagrange multiplier of the constraint $\partial_s\beta = 0$; in the
numerical method of Section~\ref{sec:numerics} it is realized, element by element, as
the dual variable of the ADMM splitting, a correspondence we exploit
both for detecting nucleation ($\max\abs{q}/q_c \to 1$) and for the
a posteriori certificates of Section~\ref{sec:num-cert}.

\subsection{The constrained-shear identity and the size effect}
\label{sec:crit-shear}

The criterion contains the classical energetic thresholds of continuum
dislocation theory as the simplest special case. Consider a single
crystal sheared between impenetrable boundaries a distance $h$ apart
along the slip direction, so that every chord is pinned--pinned of
length $h$, and let the resolved stress be uniform, $\tau = \mu\gamma$.
Then $Q(s)=\mu\gamma s$ is linear,
$\operatorname{osc} Q = \mu\abs{\gamma}h$, and case (i) gives the
energetic nucleation threshold
\begin{equation}
\mu\,\gamma_{en}\, h = 2 q_c
\quad\Longleftrightarrow\quad
\gamma_{en} = \frac{2k}{b\rho_s\,h},
\label{eq:gammaen}
\end{equation}
the size effect of the energetic threshold in constrained shear \citep{berdichevsky2007dislocation,le2008analytical}.
A point-wise criterion $\tau\le\tau_c$ is size-independent and cannot produce \eqref{eq:gammaen}; the integral criterion produces it in one line. The distinction is not academic, and the grain-size dependence of the
yield stress is its sharpest expression. A strength surface is a
point-wise condition on the stress; it contains no length, and
therefore predicts a size-independent strength: it misses not merely
the magnitude of the Hall--Petch effect but its existence. With
grain boundaries blocking the slip---the natural realization of the
condition $\beta=0$---every slip chord is pinned at both ends with
length of the order of the grain size $d$, and case~(i) of the chord
criterion applied to a uniform resolved stress gives the energetic
threshold
\begin{equation}
\tau_y \sim \frac{2 q_c}{d} = \frac{2\mu k}{b\rho_s\, d},
\label{eq:hallpetch}
\end{equation}
which means smaller grains are stronger, because the criterion integrates the
stress over the chord and the chord is the grain. The exponent carries physical information. The classical Hall--Petch relation \citep{hall1951deformation,petch1953cleavage}
scales as $d^{-1/2}$ and reflects yield governed by the transmission of slip through the boundary under the stress concentration of the dislocation pile-up; at smaller grain sizes, however, a growing body of experimental evidence supports the scaling $d^{-1}$ \citep{li1963petch,conrad1963effect,dunstan2014grain}, which is precisely \eqref{eq:hallpetch}: in small grains the energetic cost of placing
the geometrically necessary dislocations in the grain controls the
onset of slip directly, without pile-up amplification. A kindred $d^{-1}$ effect on the initial yield stress was obtained from the self-energy of geometrically necessary dislocations within strain gradient plasticity \citep{ohno2007higher}; there the linear defect energy modifies a flow rule, whereas here it enters a stability threshold of integral form, evaluated in closed form along the chords and coupled, below, to the crack. The present theory thus assigns the two observed regimes to two mechanisms---the energetic nucleation threshold at small $d$, derived here, and pile-up--mediated transmission at large $d$, which involves the dissipative threshold of the sequel and the wall of geometrically necessary dislocations that the model already forms at the pinned chord end---and predicts a crossover between them. No point-wise strength condition can produce either regime.

\subsection{Corollary: free surfaces as dislocation sources}
\label{sec:crit-surface}

Case (iii) and its degenerate limits have a physical content that we
state separately because it controls the near-threshold behavior of
notched bodies. First, on any chord meeting the free boundary at both
ends, the chord-constant slip mode is energetically free: dislocations
enter through one surface, traverse the crystal, and exit through the
other, leaving uniform slip and no stored defect content. In the
quasi-static energetic model this \emph{transit slip} proceeds at
whatever rate maintains the chord balance $\int\tau\dd{s}=0$; it
shields the stress field but stores no dislocations and, by the
argument of Section~\ref{sec:coupled}, dissipates nothing. Second, where the slip
direction is \emph{tangent} to a free surface---the grazing
characteristic through the tangency point of the chord family---the
end conditions in \eqref{eq:qcond} degenerate: trial slips concentrated
along the grazing chord achieve an unbounded ratio of extracted work to
BV cost, the threshold vanishes, and the multiplier $q$ acquires a
component concentrated on the grazing line (a measure). Physically, a
bare surface tangent to the slip plane acts as an inexhaustible
dislocation source of vanishing activation threshold; passivating the
surface ($\beta=0$ on $\Gamma_p$) removes the free ends and restores a
finite threshold given by cases (i)--(ii). Both signatures are observed
numerically and independently in Sections~5--6: the penalized
multiplier diverges as $\sqrt{C}$ along the grazing chord of the
notched disk, and the nonsmooth solver activates surface microslip
there at loads two orders of magnitude below the band-nucleation
threshold, while the passivated configuration exhibits the sharp
finite threshold of case (ii). A rigorous analysis of the grazing limit
is beyond the scope of this paper; we record it as a well-characterized
prediction of the model.

\section{Numerical method}
\label{sec:numerics}


\subsection{Incremental problem and alternate minimization}
\label{sec:num-stag}

The loading is applied as a sequence of boundary data
(Section~\ref{sec:model-bc}), and at each step the state is obtained by
minimizing the functional \eqref{eq:J} over all kinematically
admissible fields subject to the irreversibility constraint
\begin{equation}
\alpha_{n-1} \le \alpha \le 1
\quad \text{point-wise},
\label{eq:irrev}
\end{equation}
where $\alpha_{n-1}$ is the damage of the previous step; the slip
$\beta$ carries no such constraint in the present, purely energetic
model. The functional is convex in the pair $(\vb{u},\beta)$ at fixed
$\alpha$ and convex in $\alpha$ at fixed $(\vb{u},\beta)$, but not
jointly convex, and we minimize it by the alternate scheme standard in
phase-field fracture \citep{bourdin2000numerical}: Block~A minimizes
over $(\vb{u},\beta)$ at frozen damage, Block~B over $\alpha$ at
frozen $(\vb{u},\beta)$, iterated to stagnation of the energy. Each
half-step decreases $J$; this monotonicity is asserted at run time,
and any violation aborts the computation. The two blocks are convex
problems with unique minimizers, solved as follows.

\subsection{Block A: splitting and dual ascent for the slip}
\label{sec:num-admm}

At frozen $\alpha$, Block~A is a convex but nonsmooth problem: the
bounded-variation term $q_c\abs{\partial_s\beta}$ is not
differentiable at $\partial_s\beta=0$, and it is exactly this
nonsmoothness that carries the nucleation threshold of
Section~\ref{sec:criterion}; it must be treated exactly, not smoothed.
We introduce the slip gradient as an independent field,
$\xi=\partial_s\beta$, and minimize the augmented Lagrangian
\begin{equation}
L_r[\vb{u},\beta,\xi,y]
= F[\vb{u},\beta]
+ \int_\Omega q_c \abs{\xi}\dd{a}
+ \int_\Omega y\,\big(\partial_s\beta-\xi\big)\dd{a}
+ \frac{r}{2}\int_\Omega \big(\partial_s\beta-\xi\big)^2\dd{a},
\label{eq:admm-L}
\end{equation}
where $F$ collects the smooth part (elastic energy and the
$c_2$-term), $y$ is the multiplier of the constraint
$\xi=\partial_s\beta$, and $r>0$ a penalty. One iteration of the
alternating direction method of multipliers (ADMM)
\citep{boyd2011distributed} consists of
\begin{equation}
\begin{split}
(\vb{u},\beta) &\leftarrow \arg\min L_r
\quad\text{(a symmetric positive definite linear solve)},\\
\xi &\leftarrow
\mathcal{S}_{q_c/r}\!\big(\partial_s\beta + y/r\big),
\qquad
\mathcal{S}_t(z) = \operatorname{sign}(z)\max\big(\abs{z}-t,\,0\big),\\
y &\leftarrow y + r\,\big(\partial_s\beta-\xi\big),
\end{split}
\label{eq:admm-it}
\end{equation}
where the soft-threshold $\mathcal{S}_{q_c/r}$ solves the
$\xi$-subproblem point-wise and in closed form. The penalty $r$ is
selected by residual balancing during the first load steps and then
frozen; its value affects only the convergence path, not the limit.

The multiplier $y$ is the discrete realization of the micro-stress $q$
of Section~\ref{sec:crit-dual}: at convergence, $\abs{y}\le q_c$
wherever $\xi=0$, and $y=\pm q_c$ exactly on the slipped set, by the
optimality of the threshold operation. This identification is used
throughout: the approach of $\max\abs{y}/q_c$ to unity detects
incipient nucleation, and the multiplier enters the a posteriori
certificate of Section~\ref{sec:num-cert}.

\subsection{Block B: primal--dual active set for the damage}
\label{sec:num-pdas}

At frozen $(\vb{u},\beta)$, Block~B is a quadratic minimization in
$\alpha$ under the bound constraints \eqref{eq:irrev}. It is solved by
the primal--dual active-set method
\citep{hintermuller2002primal}: the sets on which the lower and
upper bounds are active are updated from the sign of the constraint
residuals, the equality-constrained quadratic problem is solved on the
inactive set, and the iteration terminates, typically in a few steps,
when the active sets stabilize. Irreversibility is thereby enforced
exactly, not by penalty or history-field surrogates.

\subsection{Discretization and meshing}
\label{sec:num-mesh}

The notched disk is discretized by linear triangles on an unstructured
mesh; the slip gradient $\partial_s\beta$ is then constant on each
element, so the auxiliary field $\xi$ and the multiplier $y$ live
element-wise and the threshold operation in \eqref{eq:admm-it} is a
single scalar operation per element. (The rectangular specimen of
Section~\ref{sec:validation} uses bilinear quadrilaterals with the corresponding
Gauss-point fields.) Mesh generation is confined behind a single
interface returning plain node and connectivity arrays, and the mesh
is refined toward the notch tip, along the expected slip rays, and
along the crack path. Crucially, the achieved element size on the
crack path is \emph{measured and asserted} at run time against the
requirement $h\lesssim \ell/4$: a computation on an unresolved mesh
aborts rather than returning plausible but wrong effective
toughnesses. The necessity of this assertion is documented, with
numbers, in Section~\ref{sec:validation}.

\subsection{Loading protocols}
\label{sec:num-load}

Two protocols are used. For nucleation measurements, the amplitude
$\kappa$ of the boundary field \eqref{eq:Kfield} is ramped
proportionally at fixed center. For toughness measurements, a crack
driven by a fixed-amplitude field with a \emph{translating} center---the surfing protocol of \citet{hossain2014effective}---reaches a
steady state in which the tip follows the moving load at constant lag,
and the fracture resistance is measured as the slope of the dissipated
energy with respect to crack advance in that steady state, free of the
stability ambiguities of fixed-center loading. The drive must be
calibrated: it has to exceed the (a priori unknown) resistance for the
crack to follow at all, yet a large overdrive with the tip
\emph{ahead} of the load center places the near field of the load on
the freshly formed wake and re-damages it, inflating the measured
resistance. We therefore choose the drive a modest margin above the
resistance measured in a first pass, and verify a posteriori that the
lag is positive; the quantitative effect of violating this rule is
reported in Section~\ref{sec:validation}.

\subsection{Verification certificates}
\label{sec:num-cert}

Every quantitative claim of Sections~5--6 is backed by run-time or
a posteriori checks that we state here because two of them do not
appear to be standard.

\emph{Equipartition as a mesh diagnostic.} For the optimal damage
profile of the functional \eqref{eq:psic}, the local and gradient
contributions to the fracture energy are equal. The ratio of the two,
computed from the discrete solution in the wake of a propagated crack,
therefore measures the adequacy of the resolution from the solution
alone: on an unresolved mesh it is large (values above~$3$ were
observed on a mesh with $h/\ell\approx 2.8$), while on resolved meshes
it saturates near $1.4$ rather than $1$---the residual excess being a
property of the irreversibly frozen two-dimensional tip profile, not
of the discretization. The ratio is thus a sensitive detector of gross
under-resolution, with a known floor.

\emph{A-posteriori primal-gap certificate.} For minimizers containing
measure-valued slip---which occur on grazing characteristics,
Section~\ref{sec:crit-surface}---the composite residual of the
iteration \eqref{eq:admm-it} can stagnate through chatter of the dual
variable on the singular set while the energetically relevant
constraint is satisfied to high accuracy. The correct certificate is
then not the solver residual but the direct energy bound: with
$\gamma = \partial_s\beta-\xi$ the primal gap of the converged state,
the quantity $q_c\int_\Omega\abs{\gamma}\dd{a}$ bounds the
contribution the unresolved gap can make to any energy in the problem,
and the computed slopes are accepted only when this bound is
negligible against the fracture energy scale $G_c\,\Delta a$ of the
measurement. In the computations of Section~\ref{sec:coupled} the bound is of order
$10^{-10}$ against a scale of $10^{-5}$.

In addition, elementary hygiene is enforced mechanically: the run mode
and all mode-dependent parameters are set by a single selector echoed
in the output, results are archived under mode-tagged names, and the
energy monotonicity of Section~\ref{sec:num-stag} is asserted at every
iteration.

\section{Validation}
\label{sec:validation}


The model contains two classical limits, and we validate each against an independent analytic result before turning to the coupled physics. With dislocations suppressed it must reproduce Griffith fracture in the regularized form of Section~\ref{sec:model-energy} (Section~\ref{sec:val-griffith}); with fracture suppressed it must reproduce the dislocation nucleation threshold of the criterion of Section~\ref{sec:criterion} (Section~\ref{sec:val-nucl}). In both cases the suppression is \emph{kinematic}. For the plastic slip this is a consequence of the theory itself: under natural boundary conditions the chord-constant transit mode of Section~\ref{sec:crit-surface} carries no defect energy, so no value of the dislocation-energy factor $k$, however large, can suppress dislocation activity energetically; the Griffith limit is defined by constraining
$\beta \equiv 0$.

\subsection{The Griffith limit}
\label{sec:val-griffith}

The notched disk of Section~\ref{sec:model-bc} is loaded with $\epsilon = 0.05$, so that the root radius $\rho_{\rm tip}=2.5\times 10^{-3}$ is an order of magnitude below the damage length $\ell = 0.02$ and the notch is effectively a sharp
crack, regularized by $\ell$ alone. With $\beta\equiv 0$, the sharp-crack
Griffith load is $\kappa_c = \sqrt{E' G_c}/\mu = 0.0302$ for
$\nu = 0.30$, $G_c = 3.2\times 10^{-4}$. The crack is driven by the
surfing protocol of Section~\ref{sec:num-load}, and the effective
fracture resistance $\Gamma$ is extracted in three independent ways: a
global linear fit of the fracture energy against the crack extension;
the fracture energy contained in a fixed window of the steady wake,
divided by the window length (a direct measurement involving no fit);
and the steady-state slope of the surfing response. Table~\ref{tab:V1}
collects the results.

\begin{table}[htb]
\centering
\caption{Griffith-limit validation: effective fracture resistance
$\Gamma/G_c$ of the propagating damage band by three independent
estimators, wake equipartition ratio $E_{\rm loc}/E_{\rm grad}$
(optimal profile: $1$), band width at $\alpha>0.5$ relative to the
theoretical $4\ell\ln 2$, and mean lag of the crack tip behind the
surfing load center. The first row documents the failure mode of an
unresolved crack path and is retained deliberately.}
\label{tab:V1}
\begin{tabular}{cccccccc}
\toprule
$h/\ell$ & drive $G/G_c$ & $\Gamma/G_c$ (fit) & (wake) & (surfing) &
equip. & width & lag \\
\midrule
$2.8$   & $1.69$ & $1.68$  & $1.68$  & ---     & $3.4$  & $0$    & --- \\
$0.18$  & $1.69$ & $1.078$ & $1.072$ & $1.082$ & $1.44$ & $1.17$ & $-0.12$ \\
$0.085$ & $1.69$ & $1.067$ & $1.059$ & $1.072$ & $1.40$ & $1.16$ & $-0.12$ \\
$0.085$ & $1.20$ & $1.059$ & $1.045$ & $1.060$ & $1.37$ & $1.14$ & $+0.04$ \\
\bottomrule
\end{tabular}
\end{table}

Four observations. First, on resolved meshes the three estimators agree to within $1\%$ at every drive, and the resistance converges toward $G_c$ from above with the expected discretization overshoot of order $h/\ell$; the residual excess extrapolated to $h\to 0$, of order $2\%$, is carried by the wake profile frozen at tip passage, which irreversibility prevents from relaxing to the one-dimensional optimum---the same effect that holds the width some $15\%$ above $4\ell\ln 2$ and the equipartition ratio near $1.4$ rather than $1$. Second, the drive must respect the positive-lag rule
of Section~\ref{sec:num-load}: at $G = 1.69\,G_c$ the tip runs ahead of the load center (negative lag), the trailing near field re-damages the fresh wake, and the measured resistance is inflated by about $1.5\%$ relative to the calibrated drive $G=1.2\,G_c$. Third, the first row shows what an unresolved crack path produces: a resistance in error by $68\%$, an equipartition ratio of $3.4$, and a band with
no measurable width---large, plausible-looking, and wrong, which is why the mesh assertion of Section~\ref{sec:num-mesh} exists and why we report the equipartition ratio as a solution-based resolution diagnostic. Fourth, the baseline is robust against the solver configuration: re-measuring the finest case under a different penalty
schedule and tolerance reproduces $\Gamma/G_c$ within one percent.

\subsection{The dislocation nucleation threshold}
\label{sec:val-nucl}

The quantitative test of the criterion of Section~\ref{sec:criterion} requires a configuration in which the chord evaluation is sharp: all slip chords pinned at both ends and crossing the boundary transversally, i.e.\ case~(i) of the chord
criterion. (The notched disk is unsuitable for this purpose by the corollary of Section~\ref{sec:crit-surface}: its free or passivated notch surface carries grazing characteristics with degenerate or measure-valued micro-stress, a prediction whose numerical signatures we confirm below.) We therefore use a square single crystal of side $1$, discretized by $200\times 200$ bilinear elements, with a
stationary crack of half-length $a_0 = 0.15$ seeded at the center
through the damage field, $\alpha = e^{-d/2\ell}$ with $d$ the
distance to the crack segment. Tension is applied by prescribed
vertical displacements $\pm\bar v$ on lubricated horizontal
boundaries, $\beta = 0$ is imposed on the entire boundary, the slip
system is inclined at $\varphi = 70.53^\circ$, and
$k = 0.42$ places the threshold inside the Rice--Thomson window of
the specimen. The damage field is frozen at the seed throughout, so
that both evaluations below address the same linear elastic state:
without this precaution the crack would begin to extend at
$\bar v \approx 0.0075$, below the slip threshold, and the comparison
would be contaminated by crack growth.

The criterion is evaluated directly: one elastic solve at a reference
load, integration of the resolved stress $\tau$ along every chord,
and the exact linear scaling of Section~\ref{sec:crit-dual} yield the
predicted threshold
\begin{equation}
\bar v_{\rm nucl}^{\rm pred} = 0.00951 ,
\label{eq:v2pred}
\end{equation}
attained on the longest crack-free chord through the ligament, with
the extrema of the micro-stress at the pinned boundary: at these
parameters the predicted first event is whole-chord yield rather than
a crack-tip dipole, the tip enhancement being outweighed by chord
length. The full non-smooth solver, run on the identical
configuration, activates slip between $\bar v = 0.0098$ and
$0.0100$: the integral slip measure $\int\abs{\xi}\dd{a}$ jumps by
four orders of magnitude across one load step, as sharp an onset as a
discrete ramp can exhibit. The measured threshold exceeds the
prediction by $4\%$, within the accounting of the $O(h)$ effects
(interpolation of $\tau$, walls occupying one element at the pinned
ends, load-step quantization). Beyond the single number, the
\emph{kinetics} of the onset match the criterion landscape: the margin
of the criterion is nearly flat across the neighboring ligament
chords, so that once the critical chord yields its neighbors follow
in a narrow load window---and the computed slip indeed grows by three
orders of magnitude over the next four load steps before settling to
steady accumulation.

The notched-disk phenomena predicted by the corollary of
Section~\ref{sec:crit-surface} are observed independently by two
methods: on the free notch, the non-smooth solver activates surface
micro-slip at loads two orders of magnitude below the band scale, with
negligible integral slip; and the penalized multiplier of the linear
evaluation diverges as $\sqrt{C}$, with the divergence localized along
the grazing characteristic through the surface tangency point. Both
are signatures of the measure-valued micro-stress predicted there,
and both disappear when the chord ends are pinned.

\section{Coupled results}
\label{sec:coupled}

We now let all three fields evolve together. The configuration is the
notched disk of Section~\ref{sec:model-bc} with $\epsilon = 0.05$, the
mesh of the finest Griffith computation ($h/\ell = 0.085$), the
passivated outer boundary and the free notch surface ($\beta = 0$ on $\partial\mathcal{D}$ as the grain boundary blocking dislocations, natural condition on $\partial\mathcal{C}$), and the physical dislocation-energy factor $k = 0.03$, so that $q_c = 6\times 10^{-4}$. The crack is driven by the two-phase protocol of Table~\ref{tab:calibration}: ramped to $G \approx 3\,G_c$ for initiation, then measured by surfing at $G = 1.5\,G_c$, comfortably above the elastic resistance measured in Section~\ref{sec:val-griffith}. Every number quoted below is certified in the sense of
Section~\ref{sec:num-cert}: two runs with different penalty schedules
reproduce the resistance slope to one part in $10^{3}$ and the initiation drive exactly (Appendix~B), and the
a posteriori primal-gap bound is $4.2\times 10^{-10}$ against the
fracture-energy scale $G_c\,\Delta a \approx 2.9\times 10^{-5}$ of the
measurement.

\begin{figure}[h]
    \centering
    \includegraphics[width=0.99\textwidth]{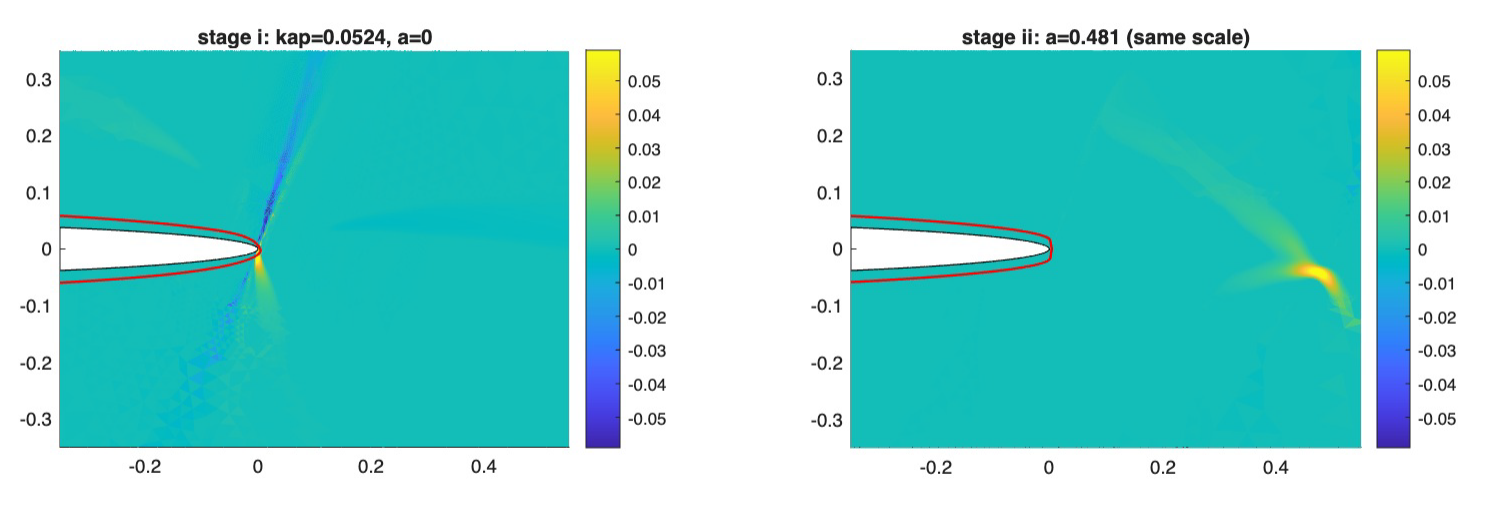}
\caption{The two stages of the coupled response: signed dislocation
density $\rho/\rho_s$ at the end of the first stage
($\kappa = 0.0524$, stationary crack, left) and in the steady
propagation of the second stage ($a = 0.481$, right). Both panels use
the identical color scale, saturated at $\pm 0.05$, chosen to
render the band interior; the maximum, $|\rho/\rho_s| \approx 2.3$, is
confined to the one-element boundary layer on the characteristic
grazing the bare notch and grows under refinement, consistent with
the measure-valued micro-stress of Section~\ref{sec:crit-surface}. Black: undeformed notch; red:
deformed notch at true scale ($\times 1$). In the first stage,
dislocation bands emanate from the notch tip and blunt it; in the
second, the bands have healed and a compact cluster of like-signed
dislocations travels with the tip at a peak standoff of
$1.4 \pm 0.1$ in units of $\ell$, its partner walls pinned at the
grain boundary (Section~\ref{sec:coupled-dipole}); the dislocation
energy of the traveling state is $0.56$ of its pre-growth value.}
    \label{fig:2}
\end{figure}

\subsection{The event sequence}
\label{sec:coupled-seq}

Slip activity begins at $\kappa = 0.0037$---an order of magnitude below the Griffith scale $\kappa_c = 0.0302$. Its onset carries the signature predicted by the corollary of Section~\ref{sec:crit-surface} for the notched geometry: the first activation is localized at the notch tip on the characteristic grazing the bare notch surface near the tip, with negligible integral slip, and the macroscopic slip content then grows continuously with the load rather than through a sharp threshold---for a notched
crystal, ``the'' emission load is an interval, not a number, exactly because the tip itself acts as a low-threshold source. Crack growth, by contrast, occurs only at the Griffith-scale drive, as in Section~\ref{sec:val-griffith}. The computed order of events---dislocation activity first, blunting next, cleavage-scale crack growth last, separated by an order of magnitude in load---is the sequence expected of a metal, produced here by energy minimization in a single functional rather than by a postulated yield condition.

\subsection{The two stages: band emission and blunting before growth; the traveling cluster during growth}
\label{sec:coupled-dipole}

\begin{figure}[h]
    \centering
    \includegraphics[width=0.5\textwidth]{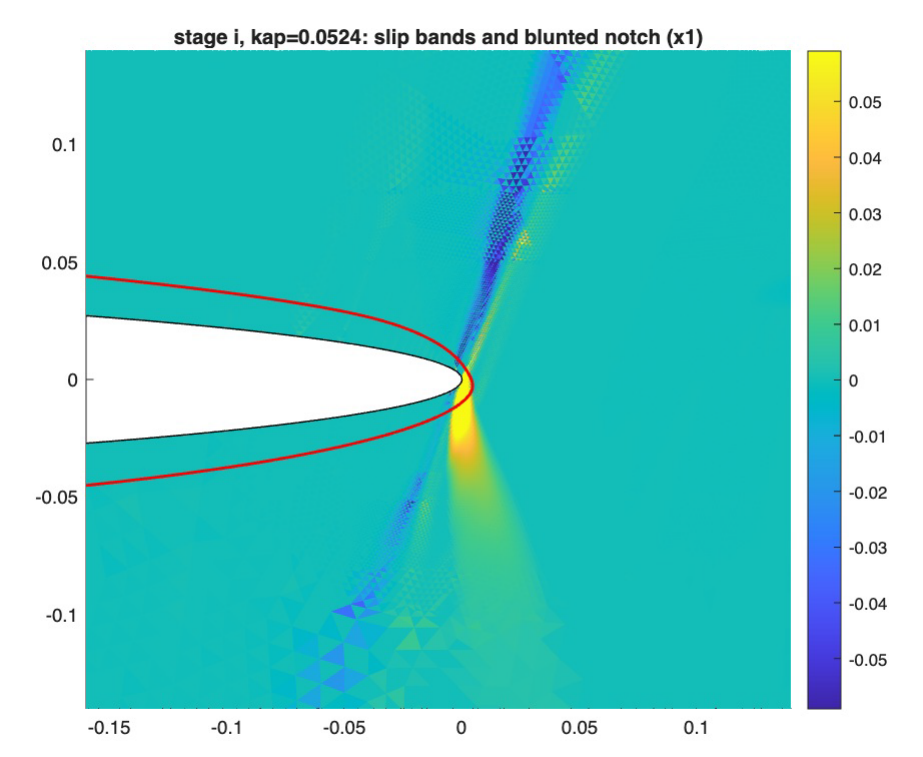}
\caption{First stage, tip region: slip bands emitted from the notch
tip at $\kappa = 0.0524$, an order of magnitude above the emission
onset and below crack growth. Signed density $\rho/\rho_s$, color
scale as in Fig.~\ref{fig:2}; the leading band meets the
bare notch surface along the grazing characteristic through the
tangency point, the low-threshold dislocation source of
Section~\ref{sec:crit-surface}. The deformed notch (red, true
scale) shows the blunting produced by the surface steps of the
bands; the undeformed profile is drawn in black.}
    \label{fig:3}
\end{figure}

The response of the coupled system divides into two stages, separated by an order of magnitude in load (Fig.~\ref{fig:2}). In the first stage, at loads beginning within the bracket established in Section~\ref{sec:coupled-seq}, dislocations are emitted in bands from the notch tip (Fig.~\ref{fig:3}). Where a band meets the free surface, the material on the two sides translates relative to itself along $\vb{s}$, producing a surface step that reshapes the notch: the tip opening displacement grows linearly with $\kappa$ in the elastic regime and departs upward from that line when slip sets in, the excess being the plastic contribution to blunting. The emitted structure shields the tip: under the loading program of Section~\ref{sec:num-load} the crack does not grow at $\kappa =1.22 \kappa_c$, and initiation is observed at $\kappa =1.73 \kappa_c$---the shielding elevates the apparent initiation load by a factor between 1.2 and 1.7 in stress intensity.

\begin{figure}[t]
    \centering
    \includegraphics[width=0.99\textwidth]{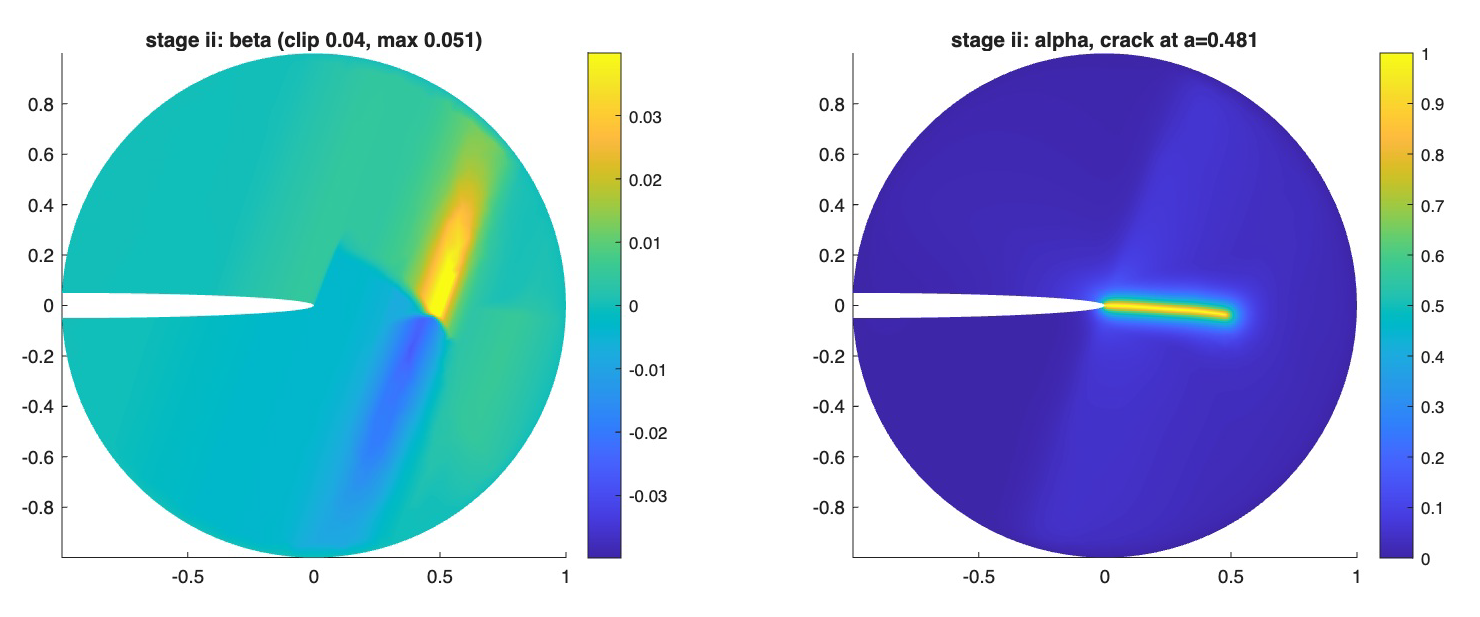}
\caption{Second stage (final state, $a = 0.481$): plastic slip
$\beta$ (left, color scale saturated at $\pm 0.04$, chosen to
render the slipped region; maximum $\beta_{\max}=0.051$ in the
grazing boundary layer) and damage $\alpha$ (right). The slip that
formed the first-stage bands has healed---$\beta$ carries no
irreversibility (Section~\ref{sec:model-energy})---and the surviving
slip is confined to the neighborhood of the current tip; the damage
field shows the crack grown from the notch along $y = 0$. Together
the two maps state the healing claim of
Section~\ref{sec:coupled-dipole} in the primal variables.}
    \label{fig:4}
\end{figure}

The second stage begins with crack growth. The band structure, carrying no irreversibility, heals as the advancing crack relieves the field that sustained it; what survives near the tip is a compact cluster of geometrically necessary dislocations of a single sign ahead of and below the crack plane---coarse-graining $N \approx 32$ discrete dislocations in $8$--$9$ walls within a region of extent $\approx 7\ell$---which travels with the tip. Its canceling partner is not local: walls of the opposite sign, $N \approx 39$, remain pinned in the boundary layer of the passivated grain boundary, so that tip cluster and boundary walls together form a grain-spanning dipole; the residual net content of the field, $-12\%$ of the total, is imaged by surface steps on the notch and crack faces where slip terminates. Measured over the steady advance 
$a\in [0.38,0.48]$, the peak standoff of the cluster from the current tip is $1.4\pm 0.1$ in units of $\ell$ ($=\ell_d$ in the present calibration; companion runs with
$\ell_d=\ell/2$ and $2\ell$ at fixed $q_c$ leave the standoff and the
cluster extent unchanged in units of $\ell$, identifying the damage
length, not the dislocation length, as the controlling scale); the drift toward larger standoff late in the window reflects the approach of the tip to the grain boundary. This is the regularized image of the Rice--Thomson emitted dislocation \citep{rice1974ductile}: the standoff is set by the competition between the attraction of the free surface and the repulsion of the crack-tip field, both encoded in the energy, and no parameter of the model prescribes it. The dislocation energy of the traveling state is 
0.56 of its pre-growth value: propagation is screened by the cluster, but the band-scale plasticity of the first stage does not survive it.

\begin{figure}[t]
    \centering
    \includegraphics[width=0.48\textwidth]{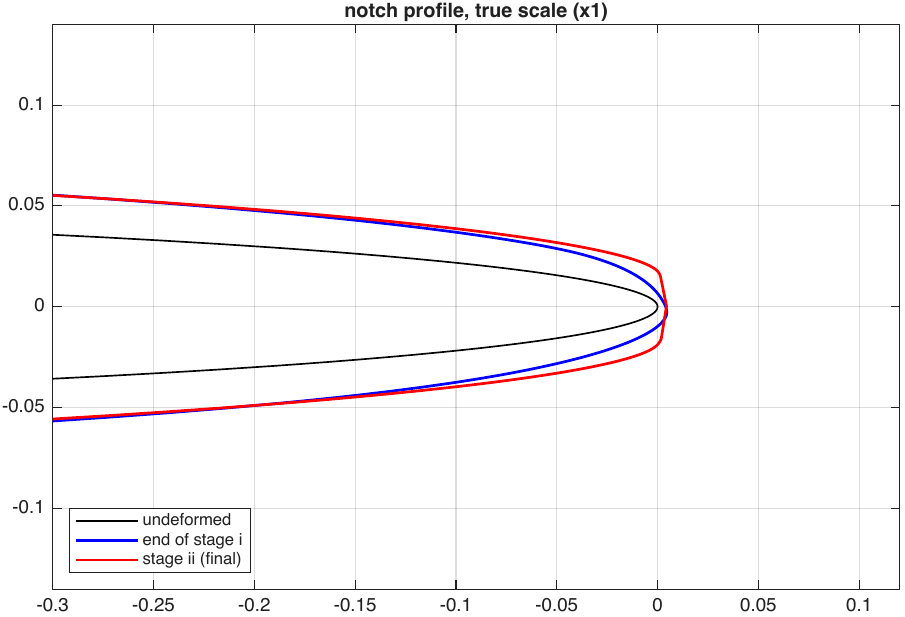}
    \includegraphics[width=0.48\textwidth]{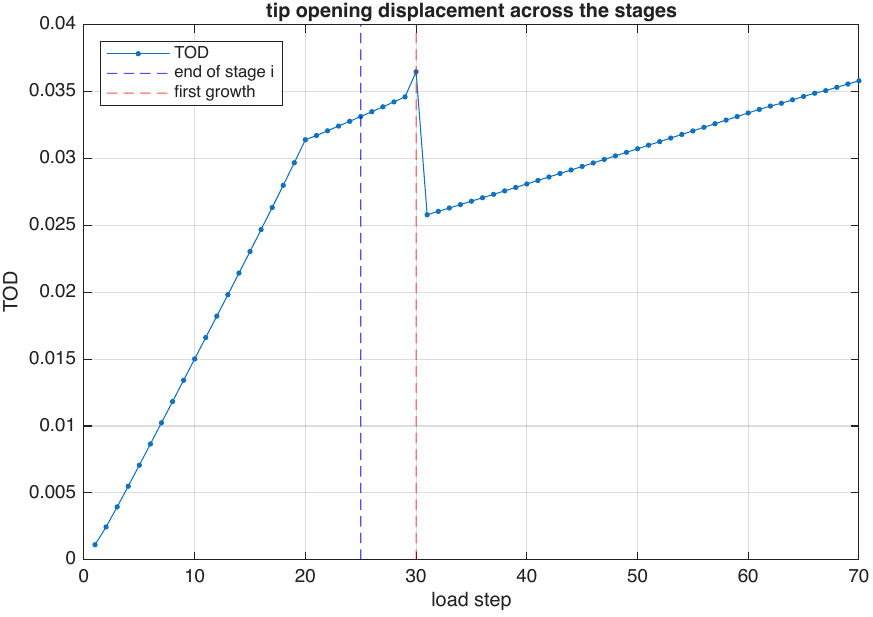}
\caption{Blunting across the two stages. Left: notch profile at true
scale ($\times 1$)---undeformed (black), at the end of the first
stage (blue), and in steady propagation (red; the load center is
then at $x_0 = 0.55$, and the opening near the original tip is
carried by the crack faces). Right: tip opening displacement over
the load program; dashed lines mark the end of the first stage
(blue) and the first crack advance (red). The departure of the TOD
from the initial linear (elastic) trend measures the plastic
contribution of the bands to blunting; its behavior at first growth
records the partial recession of the plastic blunting as the slip
heals.}
    \label{fig:5}
\end{figure}

\subsection{Zero dissipative toughening}
\label{sec:coupled-null}

The central quantitative result is stated by the decomposition of the steady-state resistance. Over the measurement window (primal-gap certificate $4.2\times 10^{-10}$ against $G_c\,\Delta a \approx 2.9\times 10^{-5}$),
\begin{equation}
\frac{1}{G_c}\,\dv{E_{\rm cr}}{a} = 1.063,
\quad
\frac{1}{G_c}\,\dv{E_{\rm dis}}{a} = 0.17,
\label{eq:null}
\end{equation}
to be compared with the purely elastic resistance $\Gamma/G_c=1.065$ of Section~\ref{sec:val-griffith} on the same mesh.
The initiation jump of the two-phase drive overshoots the translating
load center, so the tip leads the center over the early part of the
window and trails it only late (mean lag $-0.067$); the measured slope
is insensitive to this transient: restricted to the trailing-compliant
steps it is $1.064$, within one part in a thousand of the full-window
value. The two terms have different thermodynamic status. Irreversibility resides in the damage alone (Section~\ref{sec:model-energy}): the fracture term is the only dissipated energy of the model, and its slope equals the elastic resistance to two parts per thousand. The dislocation term is stored, reversible energy---recoverable upon unloading by construction---and its small positive slope is a finite-grain effect: in an unbounded domain the traveling structure would translate rigidly and 
$\dv{E_{\rm dis}}{a}$ would vanish exactly, whereas in the grain the walls pinned at the boundary and the changing chord geometry grow the stored content with advance, consistently with the standoff drift noted in Section~\ref{sec:coupled-dipole}. The value of this slope is the least certain number reported: as the small slope of a large stored content it varies by $15\%$ between the two solver schedules of Appendix B, while the resistance slope varies by one part in a thousand. Emission, screening, and blunting are all active---and the dissipated cost of fracture is unchanged.

This is the classical distinction between crack-tip shielding and toughening \citep{ritchie2011conflicts}, here derived rather than postulated, and the computation separates its two halves: shielding acts on the initiation load, elevating it by the measured factor 1.2--1.7, and screens the growing tip through the traveling cluster, while leaving the dissipated propagation resistance exactly elastic. Toughening is dissipation in the wake, and a theory without dissipation has none to offer. The formulations that couple phase-field fracture to conventional plasticity cannot pose this question, because their plastic flow is dissipative by construction \citep[e.g.][]{shanthraj2017elasto,miehe2016aphase,kristensen2020phase}; the present model isolates the energetic content of emission and shows it to be toughening-neutral. The missing ingredient is the threshold resistance to dislocation motion, which converts healing into a frozen plastic wake; its incorporation---together with the redundant dislocations and the configurational temperature of the thermodynamic dislocation theory \citep{le2018dthermodynamic}---is the subject of the sequel, for which the present result is the exact zero intercept of the toughening curve 
$\Gamma (\tau_f): \Gamma (0)=\Gamma_{\rm elastic}$, measured here to two
parts per thousand.

\section{Discussion \& outlook}
The model of this paper answers the Rice--Thomson question in its purely energetic form, and the answer has two stages. A single functional---elastic energy degraded by damage, the logarithmic dislocation energy of continuum dislocation theory, and the Ambrosio--Tortorelli fracture energy---decides by minimization alone whether the crystal creates new surface or new dislocations. The decision is quantitative and was made before the computations that confirmed it: the chord criterion predicted the nucleation threshold of the pinned square crystal to four percent, and bracketed the emission onset of the notched grain between two load programs consistently with the linear-in-$q_c$ estimate; with slip suppressed, the same functional reproduced Griffith fracture with a resistance converging to $G_c$. In the coupled computation the crystal then behaved as a metal in initiation and as a brittle solid in propagation: bands emitted an order of magnitude below the cleavage load blunted the tip and raised the initiation load by a factor 1.2--1.7; upon growth the bands healed, a compact cluster of like-signed dislocations traveled with the tip at fixed standoff---its partner walls pinned at the grain boundary---and the dissipated fracture resistance equaled the elastic one to two parts per thousand, the surviving dislocation energy being stored, not spent.

The results thus divide along a line that the model itself draws, and the division is the program of the theory. Energetics decides what nucleates and what structures exist: the threshold loads and their size dependence, the emission sites, the standoff of the cluster, the order of events. None of these requires a postulated yield condition, and none is available to a point-wise strength condition, which contains no length and therefore no size effect---the inverse-thickness threshold of constrained shear, the $d^{-1}$ yield stress of small grains, and the vanishing threshold of the grazing chord at a bare surface are all consequences of the integral character of the criterion. What energetics cannot decide is what fracture costs. The certified null of Section~\ref{sec:coupled-null} is the exact statement: in the dissipationless limit, the emitted structure translates with the tip, the slip heals in the wake, screening redistributes stored energy without altering the balance per unit advance, and $\Gamma_{\rm total} = \Gamma_{\rm elastic}$ to two parts per thousand. Emission shields---it delays initiation and screens the growing tip; it does not toughen. The classical distinction between crack-tip shielding and toughening \citep{ritchie2011conflicts} is here derived rather than postulated, and it could not have been posed within formulations that couple the damage field to conventional plasticity, whose flow is dissipative by construction and whose every increment of slip is billed to the fracture resistance whether or not it survives in the wake.

The idealizations of the paper are deliberate, and lifting them in a fixed order is the sequel. The nearest is dissipation. Endowing the slip rate with a threshold resistance $\tau_f |\dot{\beta}|$---the same degree-one structure, now in the rate, treated by the same shrinkage machinery---converts healing into a frozen plastic wake and the free transit slip of Section~\ref{sec:crit-surface} into an activated process. The wake then stores dislocation energy that does not translate with the tip, the resistance acquires an R-curve from wake buildup, and the toughening curve $\Gamma (\tau_f)$ can be measured by the surfing protocol; the null result of the present paper is its exact zero intercept, $\Gamma(0) = \Gamma_{\rm elastic}$. On the same stage enter the redundant dislocations and the configurational temperature of the thermodynamic dislocation theory \citep{le2018dthermodynamic}, which govern the storage and the stress--strain response of the wake beyond the geometrically necessary content retained here. The stage after that is temperature: thermally activated kinetics of both slip and decohesion turn the competition between emission and cleavage into a rate- and temperature-dependent one, and the brittle-to-ductile transition becomes a computable crossover of the same functional rather than a postulated criterion. The quantitative benchmark of that stage is fixed: the transition
temperature and its rate dependence measured in tungsten
\citep{gumbsch1998controlling}, already addressed by the mean-field
theories of \citet{langer2021fracture} and \citet{le2023theory}; what the
resolved fields of the present model contribute to that meeting point
are the quantities those theories must assume---the emission
thresholds, the shielding factor, and the energetic character of the
structure the tip carries. In particular, the retained cluster is
characterized by the pair $(N,\,\text{standoff}) \approx (32,\,
1.4\ell)$: precisely the coarse-grained object that the mean-field
description parameterizes, here computed rather than assumed.

Two open points are recorded rather than resolved. Mathematically, the grazing-chord limit---the measure-valued micro-stress and the vanishing threshold at a tangent bare surface---is here a well-characterized prediction with two independent numerical signatures, but its rigorous analysis, and with it the proper function-space setting of slip that is BV along characteristics only, remains to be supplied. Physically, the crossover between the energetic $d^{-1}$ regime of the yield stress derived here and the pile-up--mediated $d^{-1/2}$ regime of the Hall--Petch relation involves the dissipative threshold of the sequel and is a natural quantitative target for it.

The model is falsifiable at every stage. It commits, with no fitted plasticity parameters beyond ($b$, $\rho_s$, $k$), to the emission load of a passivated notch, to a standoff of the emitted cluster of the order of the damage length, to the $d^{-1}$ scaling of the yield stress of small grains, and to the asymmetry between bare and passivated surfaces as dislocation sources---each accessible to micro-pillar and in-situ TEM experiments at the micron scale. Energetics decides what nucleates and what exists; dissipation decides what survives and what it costs. The present paper is the exact statement of the first half.

\appendix

\section{Chord-criterion duality}
\label{app:duality}

This appendix carries out the duality computation behind the chord
criterion of Section~3.2. Throughout, $(u,\alpha)$ are frozen, so
the resolved stress $\tau = \sigma\!:\!P$ is a fixed field, and the
question is for which $\tau$ the dislocation-free state $\beta\equiv0$
minimizes the energy \mbox{$E[\beta]$} of Eq.~(10).

\paragraph{Reduction to the chords}
The operator $\partial_s$ differentiates only along the slip
direction $\vec s$. Introduce coordinates $(s,t)$ with $s$ the
arclength along the characteristics $\vec x = \vec x_0 + s\,\vec s$
and $t$ the transverse coordinate labeling them; the intersection of
each characteristic with $\Omega$ is a union of chords, each carrying
its own end conditions. Since the area element factorizes,
$da = ds\,dt$, both sides of the stability condition~(11) decompose by
Fubini:
\begin{equation}
\int_\Omega \tau\,\delta\beta\, da
 = \int\!dt \int_{\text{chord}(t)}\! \tau\, v\, ds,
\qquad
\int_\Omega |\partial_s\,\delta\beta|\, da
 = \int\!dt \int_{\text{chord}(t)}\! |v'|\, ds,
\end{equation}
where $v(\cdot) = \delta\beta(\cdot,t)$ is the restriction of the
trial slip to the chord and no transverse gradient enters the cost.
Condition~(11) therefore holds if and only if the one-dimensional
inequality
\begin{equation}
\Bigl|\int_0^L \tau\, v\, ds\Bigr| \;\le\; q_c \int_0^L |v'|\, ds
\label{eq:1dcrit}
\end{equation}
holds on (almost) every chord, for every $v$ admissible on that
chord: $v = 0$ at pinned ends, $v$ unconstrained at free ends.

\paragraph{The one-dimensional duality}
Let $Q(s) = \int_0^s \tau\, ds'$ and let $c$ be an arbitrary
constant. Integration by parts gives, for $v$ of bounded variation,
\begin{equation}
\int_0^L \tau\, v\, ds
 = \bigl[(Q-c)\,v\bigr]_0^L - \int_0^L (Q-c)\, v'\, ds .
\label{eq:ibp}
\end{equation}
The three cases of the chord criterion follow from the disposal of
the boundary term.

\emph{(i) Pinned--pinned.} Here $v(0)=v(L)=0$ and the boundary term
vanishes for every $c$. Hence
$|\int \tau v| \le \max_s |Q(s)-c| \int |v'|$ for every $c$, and the
sharpest bound is
\begin{equation}
\sup_{\int|v'|\le 1} \int_0^L \tau\, v\, ds
 \;=\; \min_c \max_s |Q(s)-c|
 \;=\; \tfrac12 \operatorname*{osc}_{[0,L]} Q .
\end{equation}
The supremum is attained in the limit by
$v' = \delta_{s_+} - \delta_{s_-}$, i.e. $v$ the indicator of the
interval between the argmin $s_-$ and the argmax $s_+$ of $Q$ (cost
$2$, extracted work $\operatorname{osc} Q$). Stability of
$\beta\equiv0$ on the chord is therefore equivalent to
$\operatorname{osc}_{[0,L]} Q \le 2 q_c$.

\emph{(ii) Pinned--free.} Let the pinned end be $s=0$ and the free
end $s=L$. The boundary term $(Q(L)-c)\,v(L)$ survives with $v(L)$
arbitrary; it is disposed of by the forced choice $c = Q(L)$, giving
the micro-stress
\begin{equation}
q(s) = Q(s) - Q(L) = -\int_s^L \tau\, ds',
\qquad q(L) = 0,
\end{equation}
and the criterion $\max_s |\!\int_s^L \tau\, ds'| \le q_c$. The
extremal trial jumps from $0$ to $\pm1$ at the argmax of $|q|$ and
rides at constant value to the free end, where it exits at no cost.

\emph{(iii) Free--free.} Constants are admissible at zero cost, so a
finite threshold requires the chord balance $\int_0^L \tau\, ds = 0$;
if the balance fails, the chord-constant mode extracts work at any
load and the threshold is zero. Under the balance, $Q(0)=Q(L)=0$,
the choice $c=0$ annihilates both boundary terms, $q = Q$ vanishes at
both ends, and the criterion is $\max_s |Q(s)| \le q_c$.

\paragraph{The micro-stress as a multiplier}
In all cases the computed $q$ is the Lagrange multiplier of the
constraint $\partial_s\beta = 0$: it satisfies $|q|\le q_c$ on the
stable chord, $q'=\tau$ in the interior, is free (a reaction) at
pinned ends and vanishes at free ends---the natural/essential
pairing dual to that of $\beta$. In the discretization of Section~4
this object is realized element-wise as the ADMM dual variable $y$:
at convergence $|y|\le q_c$ wherever $\xi = 0$ and $y = \pm q_c$ on
the slipped set, which is the discrete counterpart of the optimality
conditions above and the basis of the nucleation detector
$\max|y|/q_c \to 1$.

\paragraph{The grazing degeneracy}
Where the slip direction is tangent to a \emph{free} portion of the
boundary, the chord family degenerates. Near the tangency point of a
convex free arc, the characteristics cut segments with both ends on
the free boundary; their balance $\int\tau\,ds = 0$ fails in general,
so by case~(iii) the threshold on these chords is zero, and trial
slips concentrated on the grazing segments achieve an unbounded ratio
of extracted work to BV cost. In the limit the multiplier acquires a
component concentrated on the grazing characteristic---a measure
rather than a function. Two numerical signatures identify this
object: the penalized multiplier of the linear evaluation diverges as
$\sqrt{C}$ with the penalty constant $C$, localized on the grazing
line, and the nonsmooth solver activates surface micro-slip there at
loads far below the band scale, with negligible integral slip
(Sections~5.2 and~6.1). Pinning the arc ($\beta = 0$ on
$\Gamma_p$) removes the free ends and restores the finite thresholds
of cases (i)--(ii). A rigorous treatment of the grazing limit---and
with it the proper function space for slips of bounded variation
along the characteristics only---is beyond the scope of this paper.

\section{Numerical certificates: definitions and failure modes
observed}
\label{app:certificates}

Every quantitative claim of Sections~5--6 is backed by one of the
following checks. We record their exact definitions, acceptance
rules, and---deliberately---the failure modes we observed when they
were violated, since the latter document why each check exists.

\paragraph{B.1\quad Mesh assertion on the crack path}
At mesh generation the achieved element size along the intended crack
path is measured, and the run aborts unless the median satisfies
$h \le \ell/4$. Failure mode (Table~1, first row): at
$h/\ell \approx 2.8$ the computation returns a plausible-looking
resistance in error by $68\%$, a damage band with no measurable
width, and an equipartition ratio of $3.4$---large, convergent in
appearance, and wrong.

\paragraph{B.2\quad Equipartition as a resolution diagnostic}
For the AT2 fracture energy~(7) the optimal one-dimensional profile
$\alpha = e^{-|d|/2\ell}$ makes the local and gradient contributions
point-wise equal. The ratio
\begin{equation}
R_{\rm eq}
 = \frac{\displaystyle\int_W G_c\,\alpha^2/(4\ell)\, da}
        {\displaystyle\int_W G_c\,\ell\,|\nabla\alpha|^2\, da},
\end{equation}
computed over a window $W$ in the steady wake of a propagated crack,
measures the adequacy of the resolution from the solution alone. On
resolved meshes it saturates in the range $1.2$--$1.4$, against $\ge 3$ when unresolved; the residual
excess is carried by the two-dimensional tip profile frozen at
passage by irreversibility, not by the discretization, and
constitutes a known floor. Values well above this floor indicate
gross under-resolution (observed: $3.4$ at $h/\ell\approx2.8$,
B.1).

\paragraph{B.3\quad A-posteriori primal-gap certificate}
Let $\gamma = \partial_s\beta - \xi$ be the constraint gap of the
converged ADMM state. Since the degree-one density is
$1$-Lipschitz, $\bigl||\partial_s\beta| - |\xi|\bigr| \le |\gamma|$,
the discrepancy that the unresolved gap can induce in any energy of
the problem is bounded by
\begin{equation}
\Delta E \;\le\; q_c \int_\Omega |\gamma|\, da
 \;+\; \frac{c_2}{2}\int_\Omega |\gamma|\,
        |\partial_s\beta + \xi|\, da ,
\end{equation}
of which the second, quadratic term is below $2\times 10^{-8}$ of
the first in all reported computations. The acceptance rule for a measured slope is
$q_c\!\int|\gamma| \ll G_c\,\Delta a$ over the measurement window.
Failure mode this certificate exists for: on grazing characteristics
(\ref{app:duality}) the micro-stress is measure-valued and
the composite ADMM residual stagnates through chatter of the dual
variable on the singular set---at tolerances that would reject the
state---while the energetically relevant constraint is satisfied to
high accuracy. The correct certificate is the energy bound, not the
solver residual. In the coupled computations of Section~\ref{sec:coupled} the bound
is $4.2\times 10^{-10}$ against the fracture-energy scale
$G_c\,\Delta a \approx 2.9\times 10^{-5}$.

\paragraph{B.4\quad Energy monotonicity of the alternate scheme}
Each half-step of the alternate minimization decreases the total
energy exactly only if the inner problems are solved exactly; with
the ADMM stopped at finite tolerance, bumps of the order of the
inner-solver slack are expected and benign. The assertion is
therefore $J_{\rm new} \le J_{\rm old} + \epsilon_J(1+|J_{\rm old}|)$
with $\epsilon_J = 10^{-7}$, calibrated to exceed the observed
ADMM-slack chatter ($\Delta J \sim 10^{-9}$ absolute at energy scale
$10^{-5}$) by two orders. Violations beyond this tolerance abort the
run.

\paragraph{B.5\quad Positive-lag rule for surfing measurements}
The fracture resistance is accepted if the crack tip trails the
translating load center over the fitted window (lag $>0$), or if the
slope fitted on the trailing-compliant subwindow reproduces the
full-window slope to within the quoted precision; the latter case
arises after the initiation jump of the two-phase drive (B.6), which
overshoots the still-translating center. Failure mode (Table~1): at drive $G = 1.69\,G_c$
the tip runs ahead of the center, the trailing near field re-damages
the fresh wake, and the measured resistance is inflated by
$\approx1.5\%$ relative to the calibrated drive $G = 1.2\,G_c$.

\paragraph{B.6\quad Two-phase drive for shielded initiation}
When dislocation shielding is strong, a drive sufficient to initiate
growth exceeds the modest overdrive required by B.5. The protocol is
then two-phase: the load is ramped and held at $G_{\rm init}$ until
the crack has advanced beyond a few $\ell$, and is thereupon reduced
to $G_{\rm meas}$ for the surfing measurement; the resistance is
fitted exclusively on the $G_{\rm meas}$ window, subject to B.5.
Each drive at which initiation fails is recorded: it is a lower
bound on the shielding, $\Delta K_{\rm shield} \ge
(\sqrt{G_{\rm init}/G_c} - 1)\,\kappa_c$. In the production
calibration ($k=0.03$) the crack shows no growth when held at
$G = 1.5\,G_c$ and initiates at $G_{\rm init} = 3.0\,G_c$, bracketing
the shielding at $1.22$--$1.73$ in stress intensity.

\paragraph{B.7\quad Reproducibility and hygiene}
The run mode and all mode-dependent parameters are set by a single
selector echoed in the output; results are archived under mode- and
parameter-tagged names together with the parameter set itself, so
that every archive is self-describing. Re-measuring the finest
Griffith case under a different penalty schedule and tolerance
reproduces $\Gamma/G_c$ within one percent
(Section~\ref{sec:val-griffith}); re-running the coupled case under a
different penalty schedule and tolerance reproduces the resistance
slope to one part in $10^{3}$ ($1.0630$ versus $1.0620$), the
initiation drive and the last stationary load exactly, and the
field-quantile observables ($c_0$, $v_b$, $\beta_{\max}$) to
$2$--$3\%$; the primal-gap bounds themselves differ between schedules,
as solver residuals should, while both satisfy the acceptance rule of
B.3 with margins above $10^{4}$.

\bibliographystyle{elsarticle-harv}
\bibliography{refs}

\end{document}